# GRAPHENE AS A NOVEL SINGLE PHOTON COUNTING OPTICAL AND IR PHOTODETECTOR

J.O.D Williams*, J.S. Lapington*, M. Roy*, I.B. Hutchinson*

*Space Research Centre, Department of Physics and Astronomy, University of Leicester,
University Road, Leicester LE17RH
Corresponding Author: jodw1@le.ac.uk

## Abstract

**Bilayer graphene has many unique optoelectronic properties [1], including a tuneable band gap, that make it possible to develop new and more efficient optical and nanoelectronic devices. We have developed a Monte Carlo simulation for a single photon counting photodetector incorporating bilayer graphene. Our results show that, conceptually it would be feasible to manufacture a single photon counting photodetector (with colour sensitivity) from bilayer graphene for use across both optical and infrared wavelengths. Our concept exploits the high carrier mobility and tuneable band gap associated with a bilayer graphene approach. This allows for low noise operation over a range of cryogenic temperatures, thereby reducing the cost of cryogens with a trade off between resolution and operating temperature. The results from this theoretical study now enable us to progress onto the manufacture of prototype photon counters at optical and IR wavelengths that may have the potential to be ground-breaking in some scientific research applications.**

## Introduction

Graphene is an allotrope of carbon, specifically arranged in a 2d hexagonal lattice structure with $sp^2$ bonded carbon atoms, that has captured the world's attention since it was first isolated in Manchester in 2004 [2]. Graphene has been described as a "wonder material" due to its unique mechanical and optoelectronic properties; for instance, graphene is the strongest known material, has extremely high carrier mobility, has an approximately constant absorption coefficient as a function of frequency and is more conductive than copper [3]. Graphene is inherently a zero-band-gap-semiconductor at the Dirac points [4], although it is possible to open a gap through nanomodulation of the sample.

In this report we aim to show theoretically that graphene has the potential to work as a single photon counting colour sensitive photodetector. Bilayer graphene is formed from two layers of hexagonally arranged carbon atoms, with two possible arrangements – AA or AB (Bernal) stacking [5]. It is possible to open a tuneable band gap in the band structure of bilayer graphene through a number of techniques, such as the controlled adsorption of water molecules [6] or hydrogen [7], through nanomodulation by applying a strain [8], or by molecular doping [9]. However, for this work we consider an approach in which a potential is applied perpendicularly to the lattice [4] [10]. This breaks the interlayer symmetry and leads to the electron energy spectrum [4] given by:

$$E^2 = \gamma_0^2 |S(k)|^2 + \frac{\gamma_1^2}{2} + \left(\frac{V}{2}\right)^2 \pm \sqrt{\left(\frac{\gamma_1^2}{2}\right)^2 + (\gamma_1^2 + V^2)\gamma_0^2|S(k)|^2}$$

with a small band gap as shown in Figure 1a [10]. Additionally $\gamma_0$=2.97eV and $\gamma_1$=0.4eV [4] are the intralayer and interlayer hopping parameters respectively (illustrated in Figure 1b), and

$$S(k) = \sum_\delta e^{ik\delta} = 2\exp\left(\frac{ik_x a}{2}\right)\cos\left(\frac{k_y a\sqrt{3}}{2}\right) + \exp(-ik_x a)$$

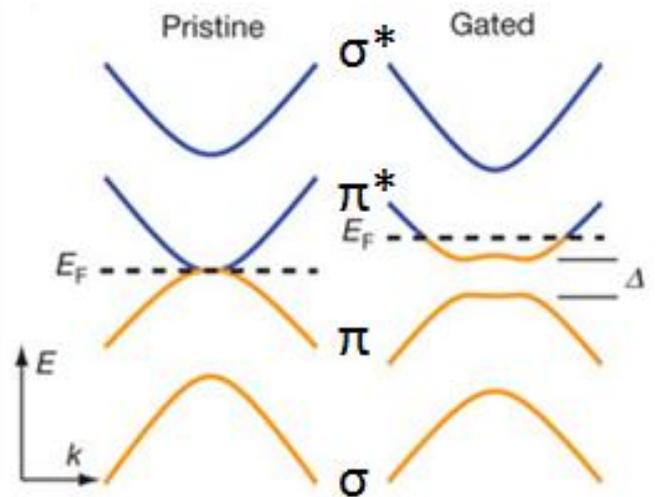

**Figure 1a.** Band structure of bilayer graphene with a band gap. Reproduced with permission of Nature Publishing Group.

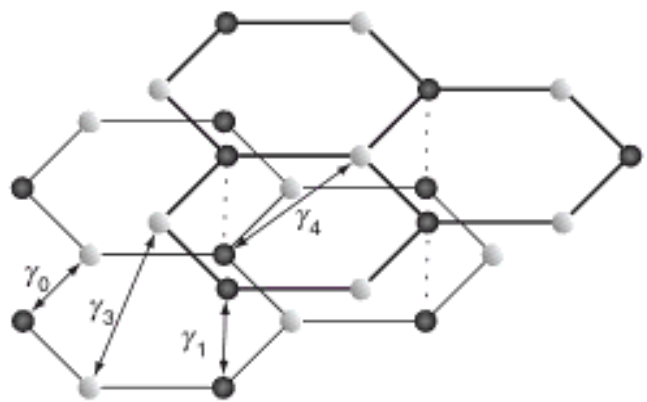

**Figure 1b.** Hopping parameters of AB stacked bilayer graphene. Reproduced by permission of Cambridge University Press subject to www.cambridge.org/uk/information/rights/permission.htm

## Density of States and Optimum Operational Window

Firstly we calculate the density of states, $n(E)$, numerically (Figure 2a) and integrate the Fermi-Dirac distribution over the first Brillouin zone to determine the number of charge carriers in the conduction band per unit area.

$$N = \int_0^{E_{photon}/2} dE \frac{1}{\exp\left(\frac{E}{k_b T}\right)+1} n(E)$$

The limits arise from the consideration of possible (vertical) photon excitations from the valence band to the conduction band, covering the energy regime we are interested in.

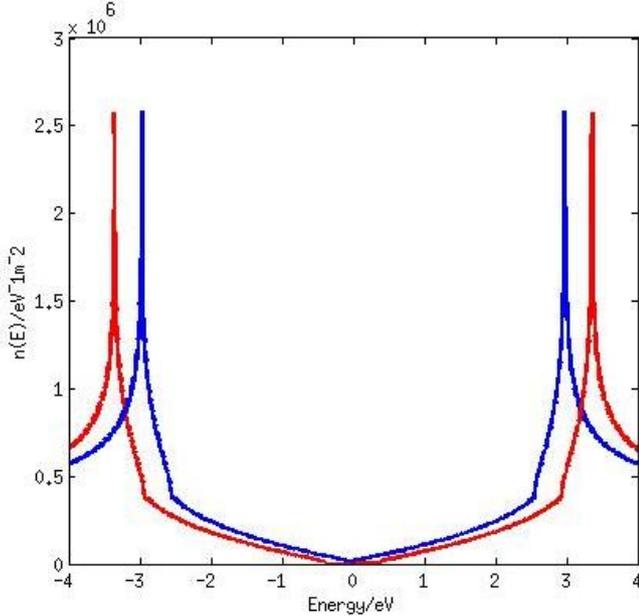

**Figure 2a.** Density of states for bilayer graphene with a band gap of 5meV. Red is the upper band, blue is the lower band.

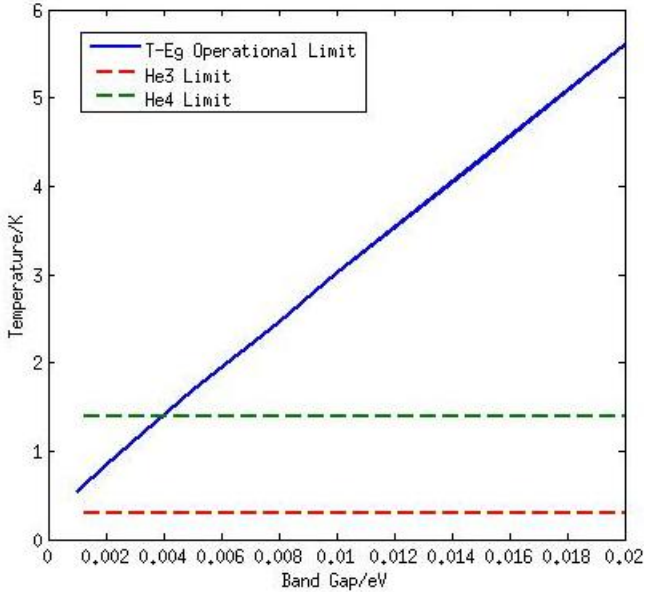

**Figure 2b.** Operational limit of a bilayer graphene photodetector. In this simulation, A=1mm².

For a single photon counting photodetector we theoretically require it to be statistically unlikely that electrons are thermally excited into the conduction band. We therefore calculate numerically NA, where $A$ is the sample area, and look for where $NA = 1$ as plotted in Figure 2b. Below this line, $NA < 1$ is the regime where there is theoretically no dark current. This is critically dependent on the bilayer graphene density of states. The tuneable band gap in bilayer graphene allows us to exploit this operational limit, as this approach allows us to run our device at higher temperatures with a larger band gap but with a trade off against resolution.

## Model and Results

In this report we employ a Monte Carlo simulation using a Gillespie Algorithm [11] to simulate the absorption of an incident photon on the graphene lattice, the excitation of a photoelectron and its subsequent relaxation in the conduction band. In this model we assume that we operate within the limit shown in Figure 2b, i.e. that electrons in the conduction band result solely from the initial photoexcitation (or subsequent relaxations). Furthermore, excitation occurs when the photon energy is equal to the energy difference between two bands in the valence and conduction bands respectively, as illustrated in Figure 1a.

After initial excitation there are a number of relaxation paths that the electron could take, namely electron-electron scattering (EES), electron-phonon scattering (EPS), impact ionisation (II) and Auger recombination (AR). EES is the inelastic scattering between two electrons in the conduction band (CB) and does not affect the total energy or the number of electrons in the CB. EPS is the scattering of an electron due to the emission (absorption) of a phonon to (from) the lattice [12], with the energy lost (gained) from the system entirely. II is the excitation of an electron from the valence band (VB) to the CB due to the loss of energy from a CB electron. In this model II is the only process which results in an increase in the number of electrons in the conduction band [13] [14] [15]. AR is the reverse process, where an electron relaxes from CB to VB, when another CB electron becomes more excited. The rates of these processes are given by [16]:

$$\sigma_{Phonon} = \sigma_{Acoustic} + \sigma_{Optical}$$
$$\approx \frac{D_0^2}{\rho_m \omega_0 (\hbar v_F)^2}(E_k - \hbar\omega_0)\left[\frac{1}{e^{\frac{\hbar\omega_0}{k_B T}}-1}+1\right]$$

$$\frac{1}{\sigma_{E-E}} = \tau_{MFT} = \frac{\lambda}{v_F} = \frac{1}{2k_f}\frac{m_e}{\hbar k_f} = \frac{1}{\pi n}\frac{m_e}{2\hbar}$$

In the literature, little work has been done on the analytical II and AR rates for low CB electron density at low temperature. However, as we start with only one conduction band electron following the photoexcitation, we assume that EES, EPS and AR relaxation rates will be significantly lower than II as the former are CB density dependent, whereas II is VB density dependent. Furthermore, relaxation rates at lower energies such that electrons relax out of CB altogether are low, due to the necessity to conserve energy and momentum whilst filling vacant holes in the VB from previous electron excitations. Therefore in the low electron density, low temperature limit, II highly dominates. To run simulations we choose a ratio, μ, of phonon scattering rate to impact ionisation rate, where II highly dominates. We run simulations with each of the relaxation events chosen randomly, weighted based on the relevant rates, and solve numerically to find solutions where energy and momentum are conserved. We test the dependence of the number of charge carriers produced as a function of time, initial photon energy, band gap and μ = $\sigma_{II}/\sigma_{Phonon}$.

*Time Dependence*

We ran our first simulations over a given time, at different initial energies and different band gaps, as shown in Figure 3. The results show, as anticipated, that the number of electrons produced increase with initial energy. Additionally, as the band gap is increased the number of electrons produced is drastically reduced.

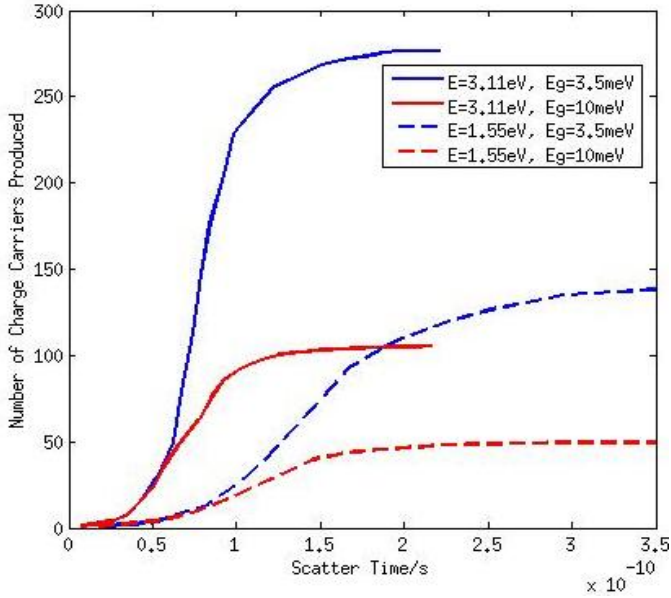

**Figure 3.** Distribution $N(t)$ at $E_{gap}$=0.01eV and 0.0035eV for photons with energy 3.11eV and 1.55eV, and with μ=100.

*Average Ionisation Energy, W*

By simulating with different size band gaps and photon energies we calculate the average electron-hole pair creation energy, $W = \frac{E_{photon}}{N}$, as shown in Figure 4a. This gives a $W$ to band gap ratio of 3-4, similar to that of semiconductors such as silicon and germanium (Figure 4b) [17].

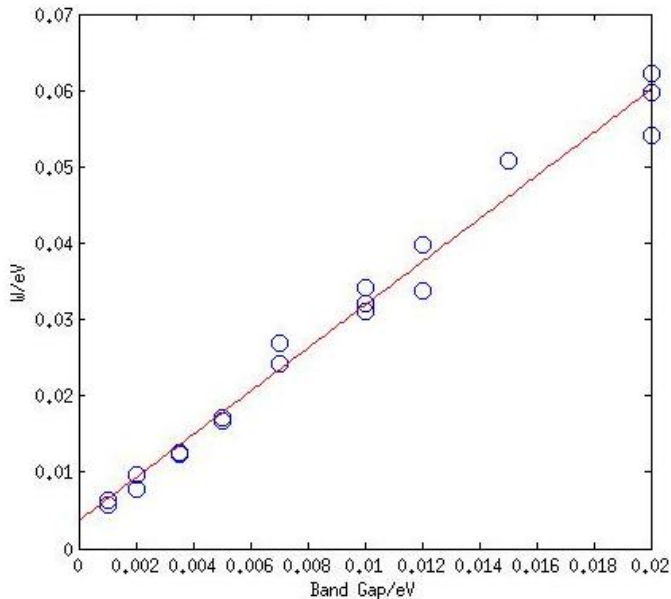

**Figure 4a.** $W$ as a function of band gap. μ=100. The circles show results from our simulations, and the red line is best fit straight line.

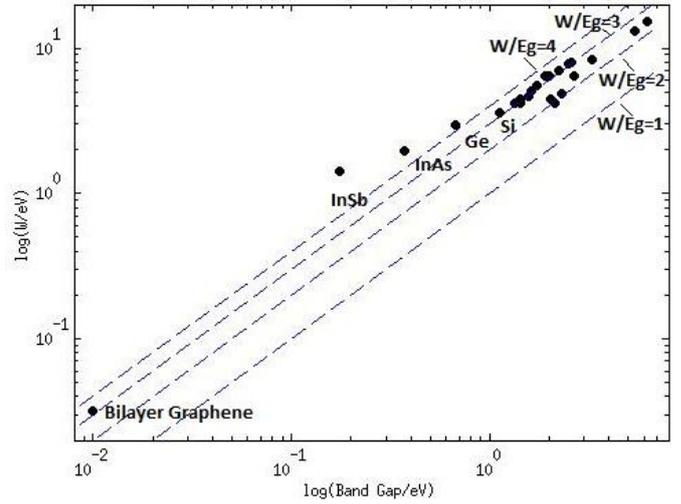

**Figure 4b.** Comparison of $W$ vs band gap for bilayer graphene with other semiconductors.

*Photon Energy Dependence*

A plot of the dependence on the initial photon energy of the distribution of charge carriers produced is shown in Figure 5a. Clearly, for a more energetic photon, more electrons will be produced proportionally to the photon energy. We observe colour sensitivity as the difference in the distributions at each energy.

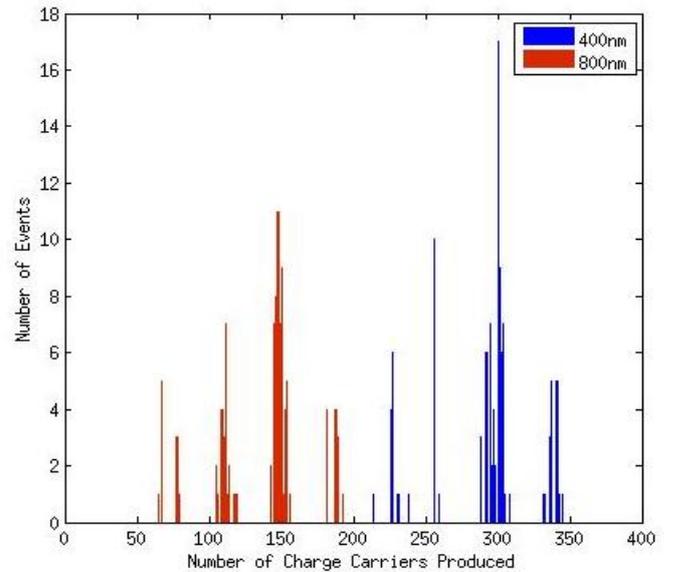

**Figure 5.** Number distribution as a function of photon energy. μ=100.

Additionally, in Figure 5 we see four peaks in the simulations, caused by the four alternative excitations from the π and σ bands to the π* and σ* bands respectively. The gap between the centre of the peaks is equal to $\Delta N = \frac{\gamma_1}{W}$, where $\gamma_1$ is the hopping parameter between layers, and $W$ is the average ionisation energy. The characteristic peak of an event is highly dependent on the initial transition between the bands, and the initial relaxation step.

As $\gamma_1$=0.4eV in our simulations, for a photon energy less than this (i.e. in the IR spectrum), we obtain only one peak. Figure 6 shows, with λ=3500nm and a band gap of 0.0035eV, that we get one large peak in the distribution, with a $W$ value still in the range, 3-4, seen in the optical.

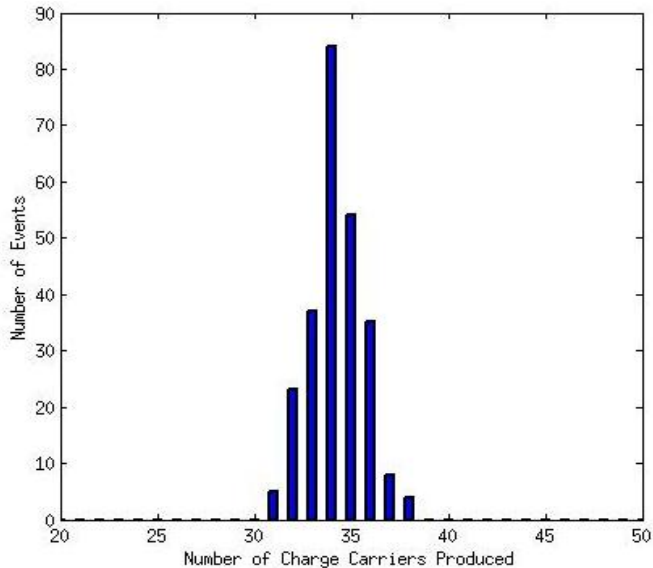

**Figure 6.** Distribution of events for λ=3500nm photon. μ=100.

*Impact Ionisation Rate Dependence*

Initially we arbitrarily picked the II rate by using a ratio to the phonon rate, $\mu$. We then tested the effect of changing the ratio to ensure that the total number of charge carriers produced tends towards the same value, but at an increased time with decreasing values of the ratio. The results are shown in Figure 7 for a photon with λ=400nm and a 0.0035eV band gap. If we integrate over the entire active scattering time (i.e. the time where electrons continue to relax and collect at the bottom of the conduction band) then this gives us an estimate of the total number of charge carriers produced. The II rate is then indicative of the active scattering time.

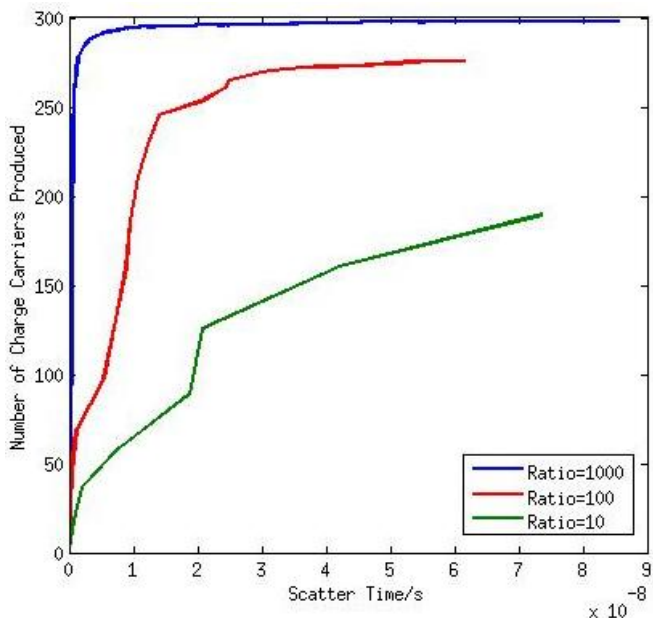

**Figure 7.** The effect of changing the II rate.

## Discussion and Conclusion

In summary, our results demonstrate the feasibility of a new type of ultrafast photon counter operating at optical and IR wavelengths with an active scattering time of order $10^{-8}$s as illustrated in Figure 6. Our concept exploits tuneable band gap which facilitates energy sensitive photon counting at selectable cryogenic operating temperatures by trading off energy resolution with operating temperature, hence allowing the use of cheaper cryogens. In a future single photon counting photodetector, we would also be able to utilise the high carrier mobility of graphene to develop an ultrafast detector.

For optical wavelengths we see fringes due to the bilayer graphene band structure, but this is not an issue for the IR wavelengths where $E_\gamma < \gamma_1$. Furthermore, the calculation of $W$ gives similar figures for those of silicon and germanium photodetectors. The variable operational temperature limit shows the flexibility offered by the tuneable band gap of bilayer graphene. This analysis gives the operating window for developing the next generation of single photon counting graphene photodetectors and our simulations allow for refinement of the model in the future.

## Acknowledgements

Simulations were performed using the SPECTRE High Performance Computer at the University of Leicester. Jamie Williams is funded by STFC.